\documentclass[prl,twocolumn,twoside,showpacs,floatfix,superscriptaddress]{revtex4-1}

\usepackage{graphicx,color,rotating}
\usepackage{amsmath}
\usepackage{dcolumn}

\usepackage{times,txfonts}

\usepackage{multirow}

\usepackage{pifont}
\usepackage{dsfont}

\usepackage{adjustbox}

\definecolor{FGViolet}{rgb}{0.61,0.32,0.61}
\definecolor{FGDarkBlue}{rgb}{0,0,0.6}
\definecolor{FGBlue}{rgb}{0,0,0.8}
\definecolor{FGLightBlue}{rgb}{0.2, 0.6, 0.8}
\definecolor{FGGreen}{rgb}{0.2,0.7,0.2}
\definecolor{FGLightGreen}{rgb}{0.4,1,0.4}
\definecolor{FGYellow}{rgb}{1,0.95,0}
\definecolor{FGOrange}{rgb}{0.95,0.5,0.1}
\definecolor{FGRed}{rgb}{0.8,0,0}
\definecolor{FGWhite}{rgb}{1,1,1}
\definecolor{FGLightGray}{rgb}{0.8,0.8,0.8}
\definecolor{FGGray}{rgb}{0.5,0.5,0.5}
\definecolor{FGDarkGray}{rgb}{0.3,0.3,0.3}
\definecolor{FGBlack}{rgb}{0,0,0}

\newcommand{\beq}{\begin{equation}}
\newcommand{\eeq}{\end{equation}}
\newcommand{\beqn}{\begin{eqnarray}}
\newcommand{\eeqn}{\end{eqnarray}}

\makeatletter
\renewcommand{\p@subsection}{}
\renewcommand{\p@subsubsection}
\makeatother

\makeatletter
\newcommand{\doublewidetilde}[1]{{%
  \mathpalette\double@widetilde{#1}%
}}
\newcommand{\double@widetilde}[2]{%
  \sbox\z@{$\m@th#1\widetilde{#2}$}%
  \ht\z@=.9\ht\z@
  \widetilde{\box\z@}%
}
\makeatother
\usepackage{titlesec}
\titlespacing\section{0pt}{24pt plus 4pt minus 2pt}{6pt plus 2pt minus 2pt}
\titlespacing\subsection{0pt}{12pt plus 4pt minus 2pt}{0pt plus 2pt minus 2pt}

\begin{document}

\title{
Fragmentation of the Giant Pairing Vibration  in $^{14}$C induced by many-body  processes
}
\author{F. Barranco}
\affiliation{Departamento de F\`isica Aplicada III,
Escuela T\'ecnica Superior de Ingenieros, Universidad de Sevilla, Camino de los Descubrimientos, 	Sevilla, Spain}	
\author{G. Potel}
\affiliation{Lawrence Livermore National Laboratory, Livermore, USA}
\author{E. Vigezzi}
\affiliation{INFN Sezione di  Milano,
Via Celoria 16, 
I-20133 Milano, Italy }

\begin{abstract}
We present a theoretical framework for treating the full excitation spectrum of $J^{\pi}=0^+$ pair addition modes, including  the well-known low-lying and bound Pairing Vibration 
on par with  the predicted Giant Pairing Vibration lying in the continuum. Our formalism  includes the coupling to low-energy collective quadrupole  modes of the core, in such a way 
 that both single-particle self-energy effects and the pairing interaction induced by  phonon exchange are accounted for. The theory  is applied to the case of  the excitation spectrum of $^{14}$C, recently populated by two-neutron transfer reactions.

\end{abstract}

\pacs{
 21.60.Jz, 
 23.40.-s, 
 26.30.-k  
 } \maketitle
\date{today}

Collective  motion is one of the distinctive features of the atomic nucleus. In particular, collective vibrations 
represent one of the most evident  manifestations of the coherent motion of neutrons and protons, displaying
transition  strengths that can exceed single-particle estimates by orders of magnitudes.  While vibrations associated with 
particle-hole (ph) and charge-exchange excitations are those most widely studied, both experimentally and theoretically, the existence of low-lying 
pairing vibrations (PV)  involving  particle-particle (pp) or hole-hole (hh) excitations has also been established  for a long time in several mass regions 
\cite{BesandBroglia66,BrogliaHansenRiedel,Brink2005} . 
Such PV  have been probed mostly by  two-nucleon transfer reactions, and the resulting cross sections and angular distributions 
 have been studied theoretically  making use of  form factors computed in the framework of the particle-particle  Random Phase Approximation (pp-RPA)  
 often neglecting ground state correlations (Tamm-Dancoff Approximation, or pp-TDA). Calculations based on the pp-TDA have been 
 often used to study pairing correlations in light and weakly bound nuclei \cite{Bertsch,Sagawa,Blanchon,Gomez}.   
 The existence of high-lying PV  (with energies of the order of 2$\hbar \Omega$,
where $\hbar \Omega \approx 41/A^{1/3}$ MeV denotes the distance between major  shells in spherical nuclei)  was proposed theoretically long ago, on the basis 
of schematic calculations within the pp-RPA \cite{BrogliaandBes1977,Bort2016}. These modes have then been studied within bound representations \cite{Herzog, Fortunato}
and taking the continuum in to account within the shell model in the Bergreen
representation  \cite{Liotta}
and  within continuum RPA calculations \cite{Khan,Avez,Matsuo}. 
While giant resonances (GR)  in the ph and charge exchange 
channels represent some  the most pronounced collective nuclear modes,   the corresponding 'Giant Pairing Vibrations'   (GPV) in the pp channel
have not been clearly identified, in spite of many experimental attempts  carried out in different mass regions and with different probes (see e.g. \cite{Assie,Laskin} and refs. therein). 
 Recent transfer experiments between heavy ions populating  $^{14}$C and $^{15}$C  \cite{Cappuzzello_2015,Cappuzzello_epj,Bonaccorso}, however,  have identified bumps 
which have been proposed to be  a signature of the GPV.

 It has been argued that one of the reasons underlying the difficulty in the identification 
of high-lying pair vibrational modes  lies in the large width that  they may acquire. It is well known that the two main mechanisms producing the damping of the giant modes obtained in the 
RPA approximation in spherical nuclei  are the coupling with more complicated configurations and the coupling with the continuum, leading to particle emission.
These effects have been  extensively studied in the case   of ph and charge exchange GR in medium mass and heavy nuclei , and it has been shown 
that including  the particle vibration coupling (PVC) with  the low-lying collective vibrations of the core  leads to considerable improvements in the calculation of the strength profile and in particular of the width, in comparison with experimental findings \cite{Niu,Litvinova,Colo}. 
Furthermore, in systems with two valence nucleons the PVC accounts for the interaction induced  
 by the exchange of  one vibrational quantum. The importance of the induced interaction on superfluidity  of infinite neutron matter is well known \cite{Schulze}.  
Such interaction renormalises the  
bare pairing interaction also in finite nuclei,  as it has been found in superfluid systems like $^{120}$Sn \cite{Schuck,Idini,Litv_Schuck}, and in light  systems like $^{11}$Li  \cite{11Li_struc}.  
The role of the continuum is particularly important in the case of  high-lying pp modes, in which case both particles involved in the excitation may be unbound.
To our knowledge, a  quantitative  microscopic calculation of the  strength function associated
with pairing modes going beyond the RPA and including continuum effects is not available. 
This is the aim of the present work. We will present results for the case of $^{14}$C,   
which are a prerequisite for a reliable computation  of absolute two-nucleon transfer cross sections.

We will consider the nucleus $A+2$  with two valence particles  on top of a vibrating   $A$  core
described by a harmonic Hamiltonian $H_{vib}$.  
The Hamiltonian of the  system A+2  is  
the sum of $H_{vib}$ and of the Hamiltonians of each  valence particle  
plus an interaction term:
\beq
                      H_{2v} = H(1)+H(2)+  V_{int}(1,2) + H_{vib},     
\label{HPV}                      
\eeq
where $H(i)= K(i) + V(i) +  H_{PVC}(i) ,i=1,2 $. $H_{vib}
= \sum_{\lambda\mu} \hbar \omega_{\lambda}  [\Gamma^{\dagger}_{\lambda\mu} \Gamma_{\lambda\mu} +1/2]$ contains the operators $\Gamma^{\dagger}$
and $\Gamma$  creating and annihilating phonons
of multipolarity $\lambda \mu$ and energy $\hbar \omega_{\lambda}$ and deformation parameter $\beta_{\lambda}$.
 The coupling between the single-particle 
motion and the core is described by $H_{PVC}$ \cite{BohrMottelson}, 
\beq
H_{PVC} (i)= \sum_{\lambda\mu} -r_i dV(i)/dr_i \; \frac{\beta_{\lambda}}{\sqrt{2\lambda+1}} 
Y_{\lambda\mu} (i) [ \Gamma^{\dagger}_{\lambda\mu} + (-1)^{\mu} \Gamma_{\lambda-\mu}].
  \label{HPVC}
\eeq
We have neglected a recoil term, that was  analyzed in detail in the case of a pp-TDA  type calculation of $^{11}$Li and found of minor relevance \cite{Esbensen1997}. 

Limiting ourselves to a first-order  calculation involving  configurations containing  at most one-phonon, 
the basis states for the A+2 system are of four different types:
 
i) fermion pp:  $| [n_{lj}n'_{lj}]_{0+}>$   ( with $e_{nlj}, e_{n'lj} > E_F)$

ii) fermion pp $\otimes$boson:  $|  [ [n_{lj}n'_{l'j'}]_{\lambda^{\pi}} \otimes  \Phi_{\lambda^{\pi}}]_{0^+}>$  ( with $e_{nlj}, e_{n'lj} > E_F)$

iii) fermion ph $\otimes$boson:    $|  [ [n_{lj}n'_{l'j'}]_{\lambda^{\pi}} \otimes  \Phi_{\lambda^{\pi}}]_{0^+}>$  ( with $e_{nlj} < E_F,  e_{n'lj} > E_F$ or $e_{n'lj} < E_F,  e_{nlj} > E_F$)

iv) fermion hh:  $| [n_{lj}n'_{lj}]_{0+}>$   ( with $e_{nlj}, e_{n'lj} < E_F)$
 
where  $E_F$ denotes  the Fermi  energy and the single particle states $e_{nlj}$
 are the eigenstates of the spherical mean field Hamiltonian, $H_{mf} = K + V ;   H_{mf} |nljm> = e_{nlj} |nljm>$. They are included 
up to an energy $E_{cut} = 30$ MeV and an orbital angular momentum $l_{cut} = $10, 
and are calculated in a spherical box of radius $R_{box}$, providing in this way a discretized continuum.
The final output of our calculation will be obtained averaging over different $R_{box}$ values, a crucial point when calculating resonant states in the continuum (see below). 
 
The Hamiltonian matrix corresponding to this extended pp-RPA calculation is schematically shown in Table \ref{Table_Ext_RPA}.
We first discuss how to determine the mean field $V$. 
It is well known that ad hoc parity- and angular momentum-dependent potentials  $V$ are often required for a reasonable 
description of single-particle states in light odd nuclei.
This is for instance  the case for the paradigmatic parity inversion occurring in $^{11}$Be \cite{Nunes}. 
In addition, one needs to make such potentials energy dependent,  in order 
to  consistently describe $0^+$ states at different energies. 
We will instead follow our previous studies of $N=7$ isotones  \cite{11Be,10Li}. There we found  that consistent  results can be obtained 
 making use of a single mean field potential, 
 determining  its few  parameters by fitting the energies of the experimental low-lying states    
including the coupling with the quadrupole degrees of freedom of the core.
In fact  the PVC  
accounts for self-energy  (including Pauli blocking) processes
which strongly  renormalize the energy of the bare single-particle states, causing the parity 
and angular momentum dependencies mentioned above, and 
naturally producing the fragmentation of their strength with associated spectroscopic factors in overall agreement with data \cite{interweaving}.
Furthermore, with our procedure we do not have to implement subtraction methods which are 
sometimes needed when collective excitations are obtained by calculations beyond mean field carried out with effective interactions previously fitted
to empirical data \cite{Tselyaev,Gambacurta}.

In the case of $^{13}$C, the resulting bare mean field potential $V(r)$, assumed to be of a Woods-Saxon shape, has a depth 
$V_0$ = 72 MeV, a diffusivity  $a = 0.65 $ fm and a radius $R =2.27 $ fm. This potential 
 produces two  deeply bound neutron hole states ($e_{1s1/2}$ = -34.0 MeV,
$e_{1p3/2}$ = -15.5 MeV). As for particle states, one finds a  $1p_{1/2}$ orbital lying at  - 5.8 MeV  and a weakly bound $2s_{1/2}$ orbital
lying at $e_{2s1/2}= $ -0.4 MeV.
Furthermore,  one finds a pronounced $d_{5/2}$  resonance at E$= 0.8$ MeV.  On the other hand, there is no low-lying $d_{3/2}$ resonance.
Including the coupling with the $2^+$ low-lying vibrational state of the core, 
using  the values $\hbar \omega_{2^+} = 4.44$ MeV and $\beta_2= 0.46$ in Eq. (\ref{HPVC})  
leads to three renormalised  many-body  particle states with energies  ${\tilde e}_{1/2^-} = -5.0,  {\tilde e}_{1/2^+} $= -2.0  and ${\tilde e}_{5/2^+}$ =  -1.4 MeV,
to be compared with the experimental energies  -4.9, -1.9 and -1.1 MeV   respectively. 
The effect of PVC on these levels is analogous to that found in $^{11}$Be (cf. Figs. 1 and 2 in \cite{11Be}). 
Note in particular the strong repulsive Pauli blocking effect on the $1/2^-$ orbital.
We do not consider renormalization effects on the hole states (namely, we ignore their coupling 
with  $ph\otimes b$ configurations, and we do not include   a $hh \otimes b $ sector in $H_{2\nu}$
 ).

We have adopted  the  finite range Gogny force for the monopole  part of the pairing interaction $V_{int}$, scaling its strength by a factor 0.9.
The action of $H_{PVC}$ induces an attractive  pairing interaction associated with the exchange of the collective modes 
between particle states. In previous studies, we have found that  experimental pairing properties are well reproduced by PVC
calculations with a reduced 
strength of the Gogny pairing force - which was determined by a fit of the empirical  odd-even mass differences -
in order to  leave room for the effects of the induced interaction \cite{Schuck}.  
We will also consider the  effects of a  quadrupole  pairing interaction acting among configurations of type-ii,
adopting a  separable interaction with a surface radial form factor, given by 
$V_{int}^{quad} = - \frac{\pi G_2}{5} \sum_{\mu} P^{\dagger}_{2\mu} P_{2\mu}$  with  
$P^{\dagger}_{2\mu} = \sum_{j_1j_2} <j_1 || dV/dr Y_2  || j_2> [a^{\dagger}_{j_1} a^{\dagger}_{j_2}]_{2\mu}$.
Notice that type-ii configurations are not renormalised within our one-phonon approximation, and we then 
use an empirical   value of the strength $G_2$ = 0.075 fm$^2$/MeV, which reproduces the estimate   
for bound states  determined in   \cite{Bes_quad}.

\begin{table}[h!]
\begin{center}
\begin{tabular}{|c|c  | c | c |c|}
\hline
$H_{2v}$ & pp & pp$\otimes b$&  ph$\otimes b$& hh   \\ \hline
 pp&  $H_{mf} + V_{int}$ & $H_{PVC}$ & $H_{PVC}$& $V_{int}$ \\ \hline
pp$\otimes$b & $H_{PVC}$ & $H_{mf}$  +  $H_{vib}$ + $V_{int}$ &   0 & 0  \\ \hline
ph$\otimes$b& $H_{PVC}$ & 0                                     & $H_{mf}$ + $H_{vib}$ + $V_{int}$  & 0 \\ \hline
hh & $V_{int}$ & 0  & 0&  $H_{mf} + V_{int}$ \\
\hline
\end{tabular}
\caption{Schematic representation of the extended pp-RPA matrix. The sub-matrix connecting pp-states  among themselves is the conventional pp-TDA matrix.
The sub-matrix connecting pp- and hh- states is the conventional pp-RPA matrix. }
\label{Table_Ext_RPA}
\end{center}
\end{table}	

We remark  that setting the deformation parameter equal to zero, 
 the usual pp-RPA for the pair addition/removal modes are recovered \cite{Blanchon,Ring}. 
We also note  that when performing standard pp-RPA, one obtains simultaneously both the pair addition (A+2) and pair removal states (A-2).
With the inclusion of states of type-iii, also A-components appear. The eigenstates 
belonging to the A+2 system are identified by selecting the eigenstates  for which the  A+2 components are the largest.
To be noted that  the value of the lowest eigenvalue  $E(0^+_1) $ corresponds to the  energy of the nucleus A+2 referred to the threshold for two particle
emission. In other words,  $- E(0^+_1) $  can be compared  to the experimental two-particle  separation energy,  $S_{2n}$  (see Fig. \ref{fig:strength_14c}(a)).


The results of the calculations  can be presented 
 by means of the  monopole pair addition strength function \cite{Matsuo_higgs}, 
 which is constructed from the N discrete eigenvalues and transition amplitudes, $E_k$ and $S_k$ ($k= 1,... N)$, 
 where the latter is computed from the eigenfunction amplitudes as 
\begin{eqnarray}
     S_k(R_{box}) =  | \sum_{pp'} X(k)_{pp'}   
    \int dr   \psi_p (r) \psi_{p'}(r) f(r) <j_p||Y_0||j_{p'}> 
       +  \nonumber \\
      \sum_{hh'}  Y(k)_{hh'} 
  \int dr \psi_h (r) \psi_{h'}(r) f(r)  <j_h||Y_0|| j_{h'}>  |^2.
  \label{strength}
\end{eqnarray}
This strength should be  representative of the relative contributions of the different spectral regions to the 
two-nucleon transfer reaction. To this purpose, we introduce a  radial form factor  $f(r)= 1/V_0 \; dV/dr$. 
The details of $S(E;R_{box})$ depend on the chosen form factor $f(r)$, but  
the position and relative strength  of collective or resonant states are quite independent of the specific form of $f(r)$,
must be concentrated in the region of the nucleus.  
 The amplitudes on the $pp-$ and $hh-$configurations are denoted by $ X(k)_{pp'}$ and $ Y(k)_{hh'}$, as in standard pp-RPA.
 
\begin{figure}[t!]
\center
\includegraphics[width=0.4\textwidth]{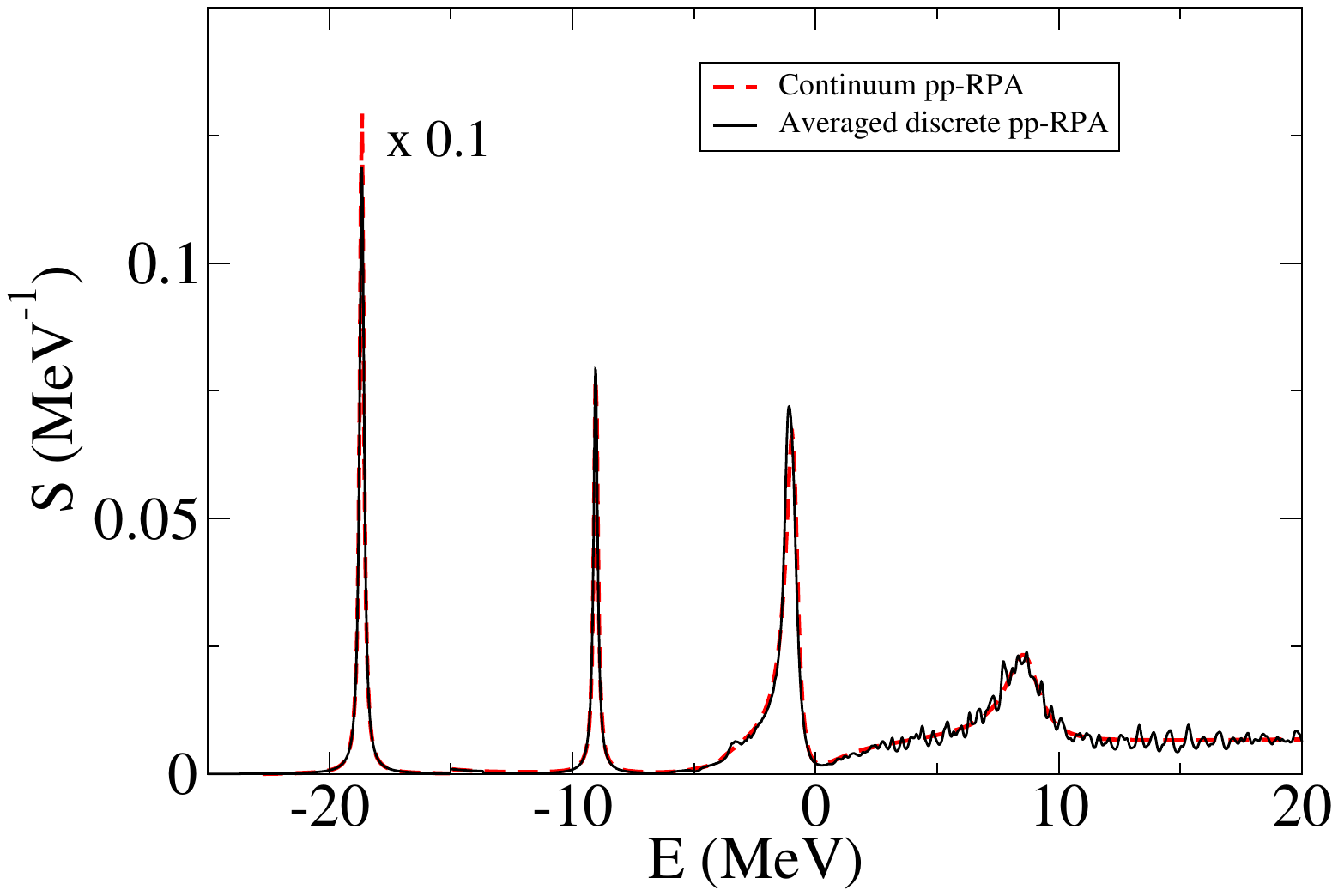}
\caption{The strength function calculated in $^{18}$O with the continuum pp-RPA is compared to the result obtained averaging the strength functions calculated in several boxes
as described in the text. The mean field has been obtained with the SLy4 interaction, while a zero-range pairing interaction
of strength $V_0$= 240 MeV fm$^3$ and an energy cutoff $E_{cut}=60$ MeV has been adopted. 
The form factor is an adimensional Fermi function, with radius $R= 1.27$ A$^{1/3}$ fm and diffusivity $a=0.65$ fm. The strength of the gs. has been reduced in the figure by a factor 0.1.
The strength functions have been averaged by a Lorentzian with FWHM =0.2 MeV. 
The continuum pp-RPA has been computed by M. Matsuo (private communication),  following the procedure 
described in \cite{Matsuo_higgs}.}
\label{fig:strength_18c}
\end{figure}

\begin{figure}[t!]
\center
\includegraphics[width=0.41\textwidth]{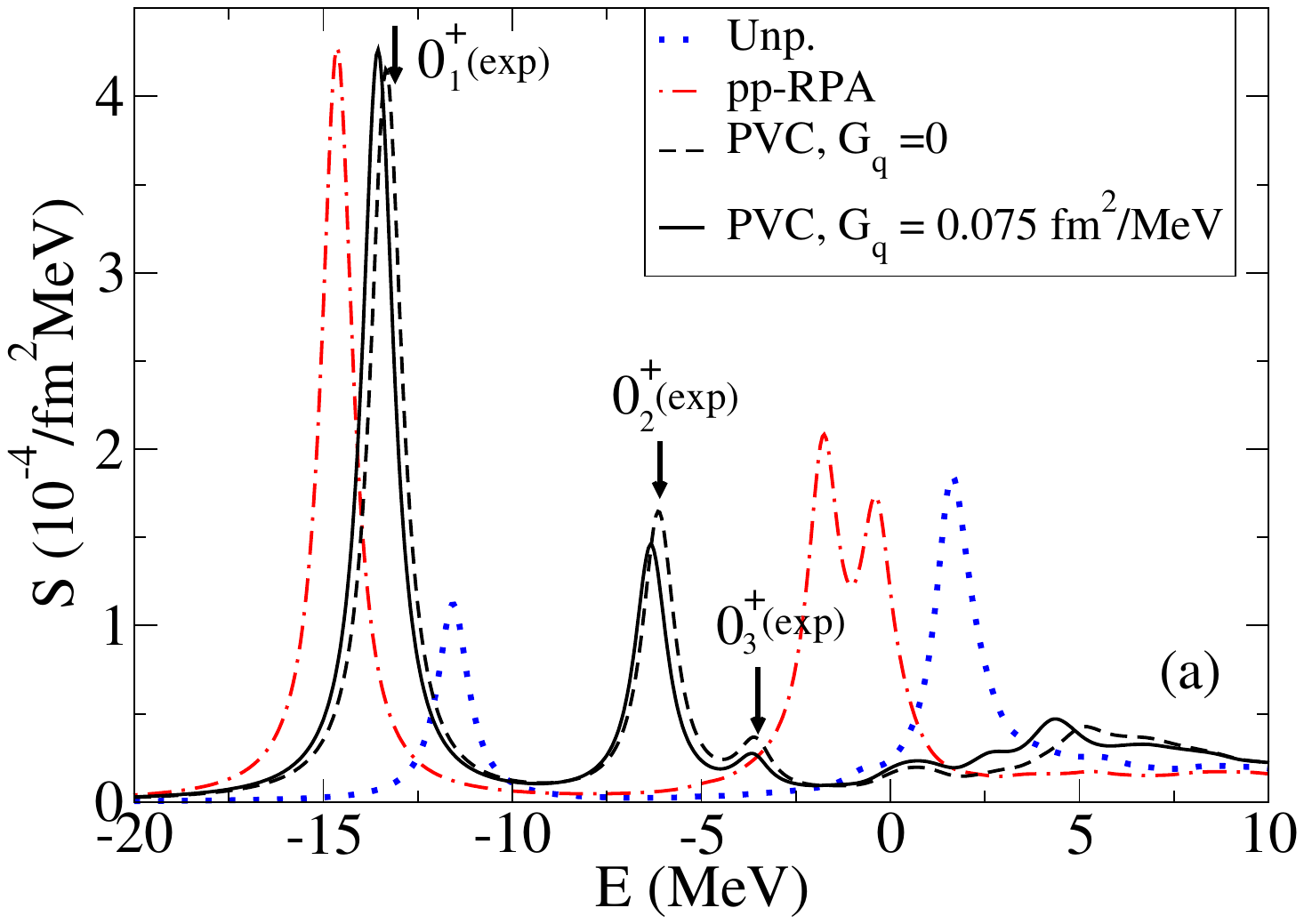}
\includegraphics[width=0.4\textwidth]{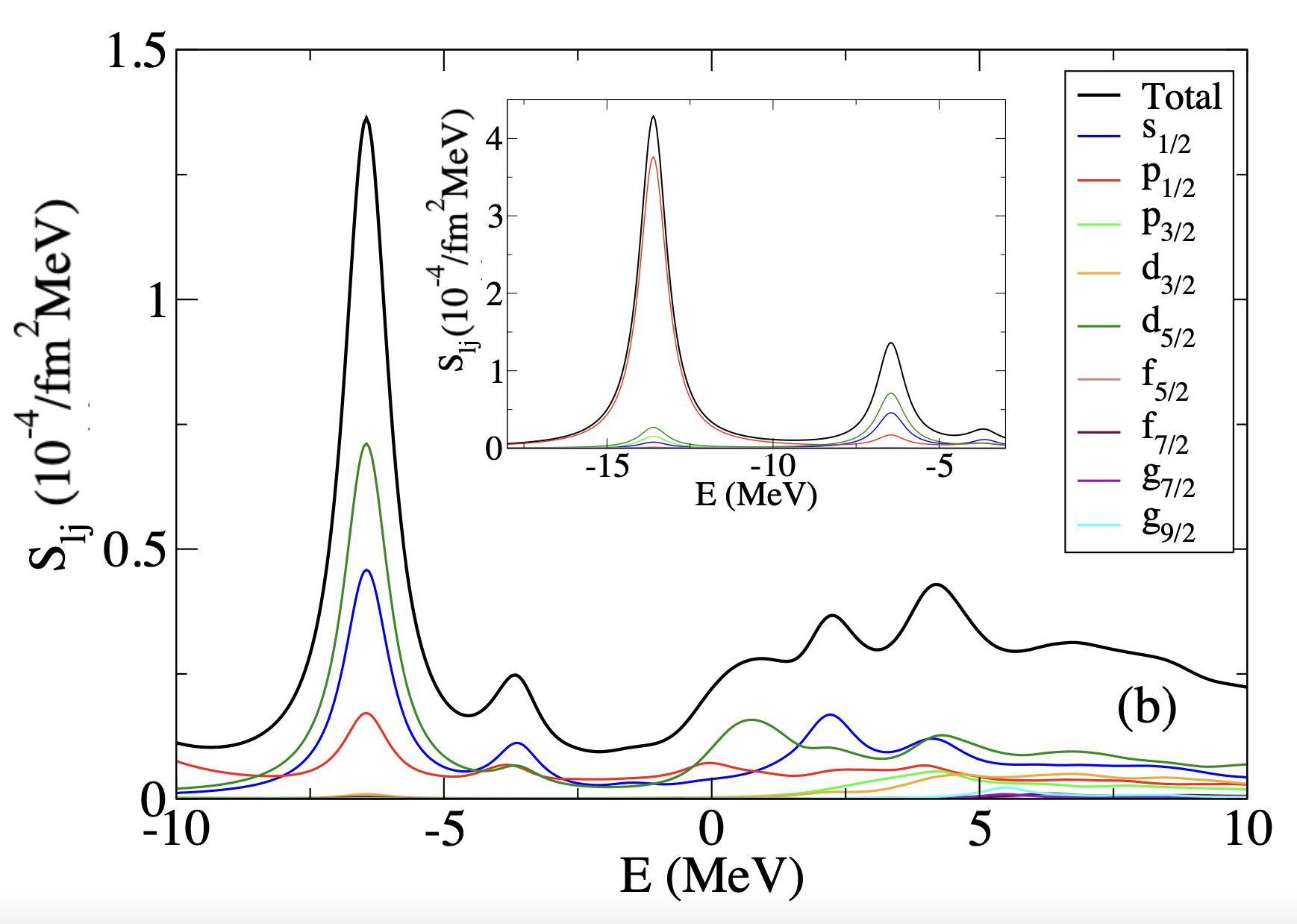}
\caption{ (a) Monopolar  strength functions in $^{14}$C, calculated as discussed in the text.
The strength functions have been averaged by a Lorentzian with FWHM =1 MeV. 
The experimental  energy $-S_{2n} $ of $^{14}$C  is indicated by the  arrow $0^+_1$.  
The positions of the two lowest excited $0^+$ states   obtained from their respective  experimental
 excitation energies  are indicated by the arrows $0^+_2$ and $0^+_3$.
(b) The PVC pairing strength function shown in (a) is displayed in the interval between $E= -10 $ and $E=+10 $ MeV in the
main plot and between $E=-18$ and $E=-3 $ MeV in the inset. The strength  is scaled with different colours according to the weight of the different angular 
momenta $lj$ in the norm $\sum_{nn'lj} [X(k)^2_{nljn'lj} - Y(k)^2_{nljn'lj}]$ of each eigenvector $|k>$.  }
\label{fig:strength_14c}
\end{figure}

  The continuous  associated strength function is obtained by taking  the convolution with a Lorentzian function $Lor(E;W)$ of convenient width $W$:
$ S(E; R_{box}) = \sum_k  S_k (R_{box}) \times  Lor(E-E_k; W)$.
In the case of  resonant states lying at positive energies, the obtained $S(E;R_{box})$ depends on the chosen value for 
$R_{box}$, the radius of the spherical reflecting wall surrounding the nucleus. 
A physically meaningful quantity is obtained  averaging
 the results obtained in a number of boxes,
$  S(E) = \frac{\sum_{i=1,N_{box}} S(E;R_{box,i})}{N_{box}}$, where $N_{box}$ is the total number of $R_{box}$ values used. 
In order for the  averaging to be meaningful, the adopted set of $R_{box}$ must satisfy certain conditions: 
the energies of the continuum levels of given angular momentum l, j should cover rather uniformly the excitation energy interval up to 
$E_{cut}$
and the energy of the lowest continuum levels  should be smaller than the energy of the resonant states under study.
In practice, we have obtained our results averaging over  the strength functions calculated in a set of 17 boxes of radius $R_{box}$ ranging from 20 fm to 28 fm 
in steps of 0.5 fm. We have validated our procedure by comparing the results obtained at the pp-RPA level (i.e., neglecting states ii) and iii)
above) with those obtained with a continuum pp-RPA calculation carried out in the nucleus $^{18}$O. 
The two strengths are almost identical,  except for small wiggles in the continuum part, as can be appreciated in Fig. \ref{fig:strength_18c}. 

\begin{figure}[h!]
\center
\includegraphics[width=0.4\textwidth]{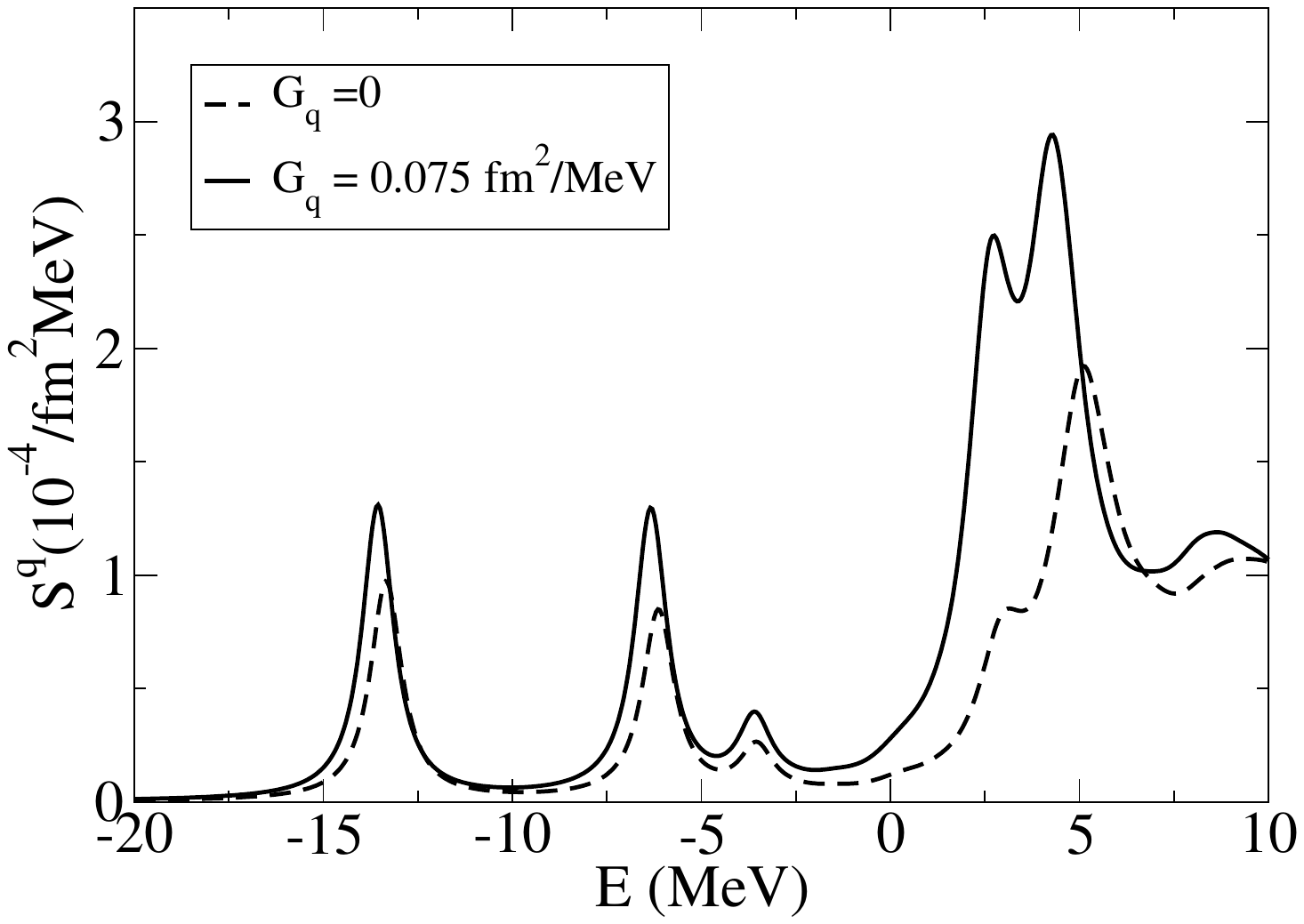}
\caption{Spectrum of the $ [ [pp']_{2^+}  \otimes 2^+]_{0^+}$ components admixed in the many-body $0^+$ states (cf. Eq. (\ref{strength_quad})).
These components result  from the diagonalization of the Hamiltonian $H_{2\nu}$ with (solid line) and without (dash-dotted line)
the quadrupole pairing interaction among the valence neutrons.} 
\label{fig:strength_quad_14c}
\end{figure}
 
 Let us now consider the $^{14}$C case.  
 In Fig.  \ref{fig:strength_14c}(a) we show the monopole strength functions  obtained with different approximations, going from the unperturbed mean field 
 calculation (curve labeled Unp) to  the pp-RPA result (pp-RPA)  and finally to the PVC strength functions obtained with  ($G_q = $ 0.075 fm$^2$/MeV) and without ($G_q$ = 0)
  the pairing quadrupole interaction.  The strength functions have been averaged with a Lorentzian curve with FWHM= 1 MeV, which is the origin of the width associated with  bound states.
  We also show in (b)  the decomposition of the full PVC result into the  contributions  $S_{lj}$ associated with the different partial waves. 
  The unperturbed strength is dominated by the lowest $(1p_{1/2})^2$ peak lying at $E \approx  -12$ MeV  
  and by the $(1d_{5/2})^2$  peak at $E \approx 2$  MeV built on the low-lying resonance. 
  The strength associated with the $(2s_{1/2})^2$ lying at $E= - 0.8$ MeV  is instead barely visible in  Fig.  \ref{fig:strength_14c}(a), because 
  the low-lying $2s_{1/2}$ orbital extends well beyond the nuclear surface and has a small overlap with the form factor.

The pp-RPA calculation lowers the energy of the ground state by about 4 MeV.
The associated pairing strength increases by a factor of about 4.
These features are those expected for a collective  PV, 
and are associated with  the presence of  a $(1d_{5/2})^2$ component (about 6\%)
and  of backward ($1p_{3/2})^2$ components  (about 4\%). In addition,
the pairing interaction shifts the  unperturbed sd bump concentrated between -1  and   +2 MeV 
down by about 4 MeV, increasing considerably its total strength. The resulting correlated bump represents the 
 GPV often discussed in more schematic models.
Comparing the pp-RPA and the PVC calculations, one can observe that  the  PVC increases the  energy of the ground state 
by almost 2 MeV, in keeping with the upward shift associated with the 1$p_{1/2}$ orbital  
discussed above in connection with  $^{13}$C. 
We remark that this is the 
main renormalization effect associated with the $ph \otimes b$  sector of our Hamiltonian (\ref{HPVC}).
The value of $E(0^+_1)$   (-13.5 MeV) is in good agreement  with the experimental two-neutron separation energy  (13.1 MeV).
 The  ground state wavefunction 
 now contains a 13\%  $pp' \otimes 2^+$ admixture, that 
 could be  tested
by the inverse transfer reaction  populating the $2^+$ state in $^{12}$C, similarly to the case of the reaction $^{11}$Li(p,t)$^9$Li$^*$ \cite{Tanihata,11Li_reac}.

 Furthermore, the PVC splits  the GPV bump described above.
 Most of the  strength is shifted to lower energy, 
 due to the  increase of  the effective mass of the  2$s_{1/2}$ and 1$d_{5/2}$ single-particle states
 discussed above in connection with $^{13}$C,  as well as to the 
effect of the induced interaction \cite{10Li}.  
As a result  one finds  two excited bound states $0^+_2$ and $0^+_3$
 lying at $E \approx$  - 6.1 MeV ($E^* \approx $ 7 MeV), with dominant $(d_{5/2})^2$ and $(s_{1/2})^2$ components, 
and  at  $E \approx -3.5 $ MeV ($E^* \approx$ 9.6 MeV), also mostly of sd character. The  $pp' \otimes 2^+$  admixtures in $0^+_2$ and $0^+_3$  
are 37\% and 33\% . 
In the recent $^{12}$C($^{18}$O, $^{16}$O)$^{14}$C transfer experiment, only the 
$0^+_3$ state was weakly populated at 
$E_{lab}$ = 84 MeV, while neither the $0^+_2$ nor the $0^+_3$ state were identified at $E_{lab}$ = 275 MeV [6, 7]. This appears to be in contrast with the present calculation, that shows a rather large strength for the $0^+_2$ state. On the other hand, these two excited states have been populated in (t,p) reactions [28] and the ratios of the measured cross sections relative to the ground state are in fair agreement with our calculated strength function. 
Considering now the GPV strength shifted to higher energy, 
the PVC produces a bump in the continuum, which is located in the excitation region $E^* \approx$ 16-20 MeV, 
not far from the bump detected in the $^{12}$C($^{18}$O, $^{16}$O)$^{14}$C experiment. 
It is found that several partial waves contribute to form this bump besides the sd orbitals.
The contribution from the $d_{3/2}$ component is not particularly significant, in keeping with the fact that the $3/2^+$ states observed  in $^{13}$C  
are either of many-body character or display a  very large width \cite{Ohnuma,Tanifuji}.
This bump has a very large admixture (about 60\%) with  $pp' \otimes  2^+$   configurations.
The energy distribution of these components, denoted by $X(k)_{pp'2^+} $, is shown in   Fig. \ref{fig:strength_quad_14c}, and is computed similarly to Eq. (\ref{strength}):
\beq
     S^q_k(R_{box}) =  | \sum_{pp'} X(k)_{pp'2^+}   
    \int dr   \psi_p (r) \psi_{p'}(r) f(r) <j_p||Y_{2}||j_{p'}> |^2
    \label{strength_quad}
\eeq
These components produce a large bump in the continuum, which   is enhanced by the action of the quadrupole pairing interaction.
One recognizes two prominent and narrow peaks.
Their width 
is much smaller than  that exhibited by the  monopolar strength shown in Fig. \ref{fig:strength_14c}, in keeping with the fact that $S^q$ involves
s- and d- waves  which are  mostly bound or resonant, because part of the excitation energy is carried by the  $2^+$ phonon.
The nature of this bump  points to the possibility to populate $0^+$ states in this region by  the combined effect of the inelastic excitation
of the 2$^+$ vibration of the core and the transfer of a pair of neutrons coupled to $2^+$. This  mechanism, would be complementary 
 to the direct two-nucleon transfer process associated with the monopole strength function,
and would also be consistent  with the  very large cross sections observed  for transfer to the $2^+$ states in $^{14}$C \cite{Cappuzzello_2015,Cappuzzello_epj}.  
Furthermore, one could argue 
about the possibility to detect quadrupole gamma transitions of the order of 4 MeV in coincidence with transfer, as a signature of the admixture of the $2^+$ vibration.

{\it Conclusions}
We have formulated an extension of the pp-RPA equations to describe $0^+$ states in the A+2 system, 
which incorporates the many-body effects associated with the dynamics of the  core. 
The theory has been applied to the pair response in $^{14}$C. 
The $^{14}$C ground state has a pronounced $(p_{1/2})^2$ character, but contains a quadrupole admixture that could be probed 
by a two-neutron transfer reaction populating the $2^+$ vibration of the $^{12}$C core.
The pp-RPA produces a concentration of sd strength close to threshold. Such strength 
represents a collective excitation in the shell next to the pair vibrational state, and is consistent with the usual  theoretical definition of   GPV.
However, the effect of PVC modifies the distribution of sd strength in an essential way. 
Most of the strength is shifted  down, producing two bound $0^+$  excited states  in reasonable agreement with experiment. 
Some structure appears in the continuum in the region where a bump was observed experimentally. 
More than 50\% of the wave functions in this region is based on components including the quadrupole excitation of the core, 
so that they may be excited by coupled channel processes. 
Challenging reaction studies are now required, in order to assess the effects of PVC on
the measured two-neutron transfer cross sections. 

{\it Acknowledgment}
This work stems from a long time collaboration with the late R.A. Broglia. We acknowledge useful discussions with G. Col\`o and M. Matsuo. F.B. acknowledges the I+D+i project
 with Ref. PID2020-114687GB-I00, funded by MCIN/AEI/10.13039/501100011033. This work was performed under the auspices of the U.S. Department of Energy by Lawrence Livermore National Laboratory under Contract DE-AC52-07NA27344.

\end{document}